\documentclass[preprint,showpacs,amsmath,amssymb,aps,prb]{revtex4}
\usepackage{graphicx}
\usepackage{dcolumn}
\usepackage{bm}
\begin{document}
\def\be{\begin{equation}}
\def\ee{\end{equation}}
\def\bea{\begin{eqnarray}}
\def\eea{\end{eqnarray}}
\title{Geometrothermodynamics}
\author{Hernando Quevedo}
\email{quevedo@nucleares.unam.mx}
\affiliation{ Instituto de Ciencias Nucleares\\
Universidad Nacional Aut\'onoma de M\'exico  \\
A.P. 70-543   \\
04510 M\'exico D.F., MEXICO}

\begin{abstract}
We present the fundamentals of  geometrothermodynamics, an
approach 
to study the properties of thermodynamic systems
in terms of differential geometric concepts. It is based, on the one
hand, upon the 
well-known contact structure of the thermodynamic phase space and, on the
other hand, on the
metric structure of the space of thermodynamic equilibrium states. 
In order to
make these two structures compatible we introduce a Legendre invariant
set of metrics in the phase space, and demand that their 
pullback generates metrics on the space of  
equilibrium states.
We show that Weinhold's metric, which was introduced {\it ad hoc}, 
is not contained within this invariant set. We  
propose alternative metrics which allow 
us to redefine the concept of thermodynamic length in an invariant manner
and to study phase transitions in terms of curvature singularities.

\end{abstract}

\maketitle

\section{Introduction}
\label{sec:int}
One of the first applications of differential
geometry in thermodynamics is due to Gibbs \cite{gibbs} and to 
Caratheodory \cite{car}.
Those studies were further developed in the seventies 
specially by Hermann \cite{her} and later by Mrugala
\cite{mru1,mru2}, leading to a geometric approach in which 
the thermodynamic phase space and its natural contact structure are
the basic elements of the construction. 
Weinhold \cite{wei1} proposed a second approach with 
a  metric  defined {\it ad hoc} as the Hessian 
of the internal energy. 
This approach has been intensively used to study 
the properties of the space generated 
by Weinhold's metric \cite{fel1,gil1}, the thermodynamic length
\cite{sal80,sal84,sal85}, the physical
properties of various two-dimensional thermodynamic systems 
\cite{nul85,san04,san05a,san05b,san05c}, and the associated 
Riemannian structure \cite{rup79,tormon93,herlac98}. 
In an attempt to understand the concept of thermodynamic length, Ruppeiner
\cite{rup79}
introduced a metric which is conformally equivalent to Weinhold's metric.  
The multiple applications of Ruppeiner's metric have been reviewed in
\cite{rup95} and more recent results are included in 
\cite{john03,jan04,san05c,shen05}.   
These two approaches have been the subject of separate analysis 
since the seventies, but the question about their relationship remains open.
Several attempts were carried out \cite{sal83,mru90}, resulting in a 
metric which was non-invariant under Legendre 
transformations. This unpleasant property means that a given
thermodynamic system has different properties, when
different thermodynamic potentials are used.

The main purpose of the present work is to show that both approaches 
can be unified into a single approach by using purely mathematical 
considerations. We will call this approach {\it geometrothermodynamics}.
The idea is simple. On the phase space we first consider 
an arbitrary metric which, by means of the pullback, induces 
a metric on the space of equilibrium states. 
 Then we derive the conditions 
for these metrics to be invariant with respect to arbitrary Legendre 
transformations, making the contact structure compatible with the 
Riemannian structure.   
Geometrothermodynamics is 
a unifying approach which allows us to treat thermodynamics in 
a geometric language either at the level of the phase
space or in the space of  equilibrium  states. 
The first attempt towards
a formulation of geometrothermodynamics was presented in \cite{quezar03}
where the connection, torsion and curvature of the
fundamental Gibbs 1-form were used to construct a 
geometry for equilibrium states.

This paper is organized as follows. In Section \ref{sec:gtd} we present the
main geometric structures which are necessary for the formulation of
geometrothermodynamics. This includes the contact 
structure, Legendre invariance, and the Riemannian structures.
In Section \ref{sec:2dof} we present the special  
case of two-dimensional thermodynamic systems. We analyze
Weinhold's approach and show that it does not satisfy all the 
requirements necessary to describe thermodynamic systems in 
an invariant manner.
We find the conditions for a metric to be Legendre invariant 
and show the existence of metrics with
this property. Furthermore,
in Section \ref{sec:ig}  we analyze the simple case of an ideal  gas
and propose a metric whose curvature can be used as a measure of
thermodynamic interaction. The same metric is used to propose an 
invariant definition of thermodynamic length. In Section \ref{sec:vdw} 
we analyze the van der Waals gas and prove that critical points
and phase transitions of any two-dimensional thermodynamic system
are characterized by curvature singularities 
of a specific metric.  
 Finally, Section \ref{sec:con} is devoted
to discussions of our results and suggestions for further research.
\section{Fundamentals of geometrothermodynamics}
\label{sec:gtd}

\subsection{Contact structure}
\label{sec:contact}

The first step to introduce the language of 
differential geometry in thermodynamics is the definition of the 
$(2n+1)$-dimensional {\it thermodynamic phase space} ${\mathcal  T}$. It
can be coordinatized  by the set $Z^A=\{\Phi, E^a, I^a\}$, where 
$\Phi$ is the thermodynamic potential, and $E^a$ and $I^a$ denote the
extensive and intensive variables, respectively. Here we adopt the conventions
$A=0,..., 2n$ and $a=1,...,n$ so that $ \Phi = Z^0$, $E^a=Z^a$, and $I^a = Z^{n+a}$.
Furthermore, we introduce the fundamental Gibbs 1-form (summation over repeated indices)
\be
 \Theta_{_G} = d\Phi - \delta_{ab} I^a d E^b \ ,\quad \delta_{ab}={\rm diag} (1,1,...,1)
\label{gibbs}
\ee
The pair
$({\mathcal T}, \Theta_{_G})$ is called a contact manifold \cite{her} 
if ${\mathcal T}$ is differentiable and $\Theta_{_G}$ satisfies the 
condition 
$\Theta_{_G} \wedge (d\Theta_{_G})^n \neq 0$.
Consider now the $n-$dimensional space ${\mathcal E}$ spanned by the coordinates $E^a$.
This can be realized by means of the smooth mapping $ \varphi : \   {\mathcal E} \  \longrightarrow  {\mathcal T}$ 
\be
{}  \varphi :  (E^a) \longmapsto  (\Phi, E^a, I^a)\ ,
\label{map}
\ee
with $\Phi=\Phi(E^a)$. 
We define the {\it space of 
thermodynamic equilibrium states} ${\mathcal E}$
 as the subspace of ${\mathcal T}$ given by
the embedding mapping $\varphi:  {\mathcal E} \rightarrow {\mathcal T}$  for which the condition 
\be
\varphi^*(\Theta_{_G})= \varphi^*\left(d\Phi - \delta_{ab} I^a d E^b\right) = 0 
\label{gibbsdown}
\ee
holds, where $\varphi^*$ represents the pullback. This implies  the relationships
\be
\frac{\partial\Phi}{\partial E^a} = \delta_{ab}I^b
\label{dual}
\ee
which are the standard conditions of thermodynamic equilibrium. 
Consequently, on the space of equilibrium states ${\mathcal E}$,
condition
(\ref{gibbsdown}) leads to the {\it first law of thermodynamics}
\be
d\Phi  - \delta_{ab} I^a d E^b = 0 \ .
\label{flaw}
\ee
To distinguish one thermodynamic system from another in the space of
equilibrium states,
 one can specify the fundamental equation \cite{callen} which,
in the representation used here, 
is contained in the embedding mapping (\ref{map}) through the relationship
$\Phi = \Phi (E^a)$.  
This construction must be complemented with the {\it second law of thermodynamics}
\be
\frac{\partial^2 \Phi}{\partial E ^a \partial E ^b} \geq 0 \ ,
\label{slaw}
\ee
which is also known as the convexity condition \cite{burke,isr79}. 
The thermodynamic potential must satisfy the homogeneity condition 
$\Phi(\lambda E^a) = \lambda^\beta \Phi(E^a)$ for constants parameters
 $\lambda$ 
and $\beta$. 
Differentiating this homogeneity condition with respect to $\lambda$ and
evaluating the result at $\lambda=1$, we get
an expression which transforms into Euler's identity
\be
\beta \Phi(E^a) = \delta_{ab}I^b E^a \ ,
\label{euler}
\ee
after using the duality condition (\ref{dual}). Calculating
the exterior derivative of Euler's identity and using (\ref{flaw}), 
we obtain the generalized Gibbs-Duhem relation 
\be
(1-\beta)\delta_{ab}I^a d E^b +\delta_{ab} E^a d I^b = 0 \ .
\label{gibduh}
\ee
The classical expressions for Euler's identity and Gibbs-Duhem relation 
are obtained from the above equations by putting $\beta = 1$.

\subsection{Legendre invariance}

Legendre transformations are a special case of contact transformations 
which leave invariant the contact structure of ${\mathcal T}$. 
In physical terms, Legendre invariance means that the
thermodynamic properties of a system are independent 
of the thermodynamic potential used to describe it. 
Let us consider a partial Legendre transformation \cite{arnold}  
\be
\{Z^A\}\longrightarrow \{\widetilde{Z}^A\}=\{\tilde \Phi, \tilde E ^a, \tilde I ^ a\}
\ee
\be
 \Phi = \tilde \Phi - \delta_{kl} \tilde E ^k \tilde I ^l \ ,\quad
 E^i = - \tilde I ^ {i}, \ \  
E^j = \tilde E ^j,\quad   
 I^{i} = \tilde E ^ i , \ \
 I^j = \tilde I ^j \ ,
 \label{leg}
\ee
where $i\cup j$ is any disjoint decomposition of the set of indices $\{1,...,n\}$,
and $k,l= 1,...,i$. In particular, for $i=\{1,...,n\}$ and $i=\emptyset$ we obtain
the total Legendre transformation and the identity, respectively. In the further
discussions we will always consider total Legendre transformations for the 
sake of simplicity. The main results will not be affected by this simplification.

The invariance of the contact
structure introduced above 
becomes evident from the fact that in the new coordinates $\{\tilde Z ^A\}$ the 
Gibbs 1-form (\ref{gibbs}) becomes 
$\tilde \Theta _{_G} = d\tilde \Phi - \delta_{ab} \tilde I^a d \tilde E^b$. 
A straightforward calculation proves also the  
invariance of the duality condition (\ref{dual}) and the 
first (\ref{flaw}) and second (\ref{slaw}) laws of thermodynamics.
Notice that a Legendre transformation interchanges the character of the 
extensive and intensive variables. For this reason, to establish the validity
of Euler's identity and, consequently, the Gibbs-Duhem relation in the new
variables it is necessary to identify first the extensive variables $\tilde E ^a$
and the fundamental equation $\tilde \Phi = \tilde \Phi (\tilde  E ^a)$. Then,
Legendre invariance follows from the fact that the duality condition 
(\ref{dual}) and the first law (\ref{flaw}) are invariant.   
\subsection{Metric structure}
\label{sec:leg}

Let us consider a non-degenerate metric $G$ which induces a Riemannian 
structure on the thermodynamic phase space
${\mathcal T}$. 
Now we can formulate the main statement of geometrothermodynamics.
A thermodynamic system  is described by  a 
metric $G$ which is called a {\it thermodynamic metric}
 if it satisfies the following conditions:

{\bf 1.} $G$ is invariant with respect to transformations which do not 
modify the contact structure of ${\mathcal T}$. In particular, $G$ must be invariant
with respect to Legendre transformations.

{\bf 2.} $G$ induces in the space of equilibrium states ${\mathcal E}$ an invariant 
metric $g$ by means of the mapping
\be
\varphi^* (G) = g \ .
\label{mapg}
\ee
The idea behind condition 1 is that a thermodynamic metric must describe a
thermodynamic system, independently of the coordinates used in ${\mathcal T}$. 
This condition is necessary in order for geometrothermodynamics
to be able to describe thermodynamic properties in terms of geometric
concepts in a manner  which must be invariant with respect to changes 
of the thermodynamic potential. Condition 2 establishes the relationship 
between the geometry on ${\mathcal T}$ and the geometry on 
${\mathcal E}$ by using the same
tools that are used to define equilibrium states 
in the contact structure approach. 

It is clear that the Legendre
invariance of $g$ follows from the invariance of $G$, if they are 
related by Eq.(\ref{mapg}). In fact, if we denote by $\tilde G = 
\tilde G (\tilde Z ^A) $
the metric obtained from $G$ by applying the Legendre transformation 
(\ref{leg}), and define $G^\prime(\tilde Z ^A) = G(Z^A = \tilde Z ^A)$, 
then the Legendre
invariance of $G$ can be reached by demanding that $\tilde G (\tilde Z ^A)
= G^\prime (\tilde Z ^A)$. Furthermore, since $g=\varphi^*(G)$ and 
$g^\prime = \tilde \varphi ^* (G^\prime)$ with $\tilde \varphi$ defined
as in Eq.(\ref{map}) with $ Z^A = \tilde Z ^A$, then we have that 
$\tilde g = \tilde \varphi^* (\tilde G) = \tilde \varphi ^* (G^\prime)
= g^\prime$, i.e. $g$ is an invariant metric.
   
Under a total Legendre transformation, $ Z^A \rightarrow \tilde Z ^A$, as given 
in Eq.(\ref{leg}),  the components of the metric $G$ transform 
as
\be
G_{AB} \rightarrow \tilde G _{AB} = \frac{\partial Z^C}{\partial \tilde Z ^A}
  \frac{\partial Z^D}{\partial \tilde Z ^B} G_{CD} \ ,
\ee
with the transformation matrix
\be
\frac{\partial Z^A}{\partial \tilde Z ^B}
=\left( \begin{array}{ccccccccc} 
1& \ - \tilde Z ^{n+1} & \ -\tilde Z ^{n+2} & ... & \ - \tilde Z ^{2n} 
&   \ - \tilde Z ^{1} & \ -\tilde Z ^{2} & ... & \ - \tilde Z ^{n}                   \\
0 & 0 & 0 & ... & 0
& -1  & 0 & ... & 0  \\
. & . & . & ... & .
& .  & . & ... & .  \\
0 & 0 & 0 & ... & 0
& 0  & 0 & ... & -1  \\
0 & 1 & 0 & ... & 0
& 0  & 0 & ... & 0  \\
. & . & . & ... & .
& .  & . & ... & .  \\
0 & 0 & 0 & ... & 1
& 0  & 0 & ... & 0  \\
\end{array}
\right) \, \,\, ,
\label{matgen}
\ee
where the indices $A$ and $B$ represent rows and columns, respectively.
Since the determinant of this matrix is equal to one, 
the inverse transformation exists, and it determines the 
diffeomorphism corresponding to a total Legendre transformation.  

To guarantee that the metric $G$ in ${\mathcal T}$ is Legendre
invariant it is sufficient to demand that each of its components preserves its functional
dependence after a Legendre transformation. Let 
$\tilde G _{AB}(\tilde Z ^C) = G_{AB}(Z^C\rightarrow \tilde Z ^C)$ represent the components 
of the metric $\tilde G$ obtained after a Legendre transformation (\ref{leg}) of the metric
$G$, and let 
$G^\prime_{AB}(\tilde Z ^C) = G_{AB}(Z^C = \tilde Z ^C)$ be the components of the metric 
$G^\prime$ obtained by replacing each coordinate $Z^C$ by its counterpart 
$\tilde Z ^C$ in the
metric $G$ (no Legendre transformation is applied). Then, if we demand that the condition
\be
\tilde G _{AB}(\tilde Z ^C) = G^\prime_{AB}(\tilde Z ^C)
\label{invcon}
\ee
holds for each value of $A$ and $B$, we guarantee that the metrics $G$ on 
${\mathcal T}$ as well as $g$  on ${\mathcal E}$ are Legendre invariant. 

An important issue to be addressed when imposing invariance conditions on metrics
is the one related to the existence of solutions. In our case, the point is whether
there exist metrics which satisfy the condition of invariance under Legendre
transformations as given, for instance, in Eq.(\ref{invcon}). 
In fact, it is easy to construct an invariant $(2n+1)-$dimensional metric by
defining the Gibbs metric as the ``square" of the Gibbs 1-form (\ref{gibbs}), i.e.
\be
G = \Theta_{_G} \otimes \Theta_{_G} = \Theta_{_G} ^2
= d\Phi ^ 2 - 2 \delta_{ab}I^a d E^b d\Phi 
+ \delta_{ab}\delta_{cd} I^a I^c d E^b d E^ d \ . 
\label{gibbsmet}
\ee
The invariance of this metric follows from the invariance of the Gibbs 1-form.
Since the Gibbs metric does not possess components proportional to $d I^a$,
its determinant vanishes. However, it contains an ($n+1)$-dimensional
metric in the sector $\{d\Phi, d E^a\}$ with non-vanishing determinant. 
Despite its degeneracy, later on we will use the Gibbs metric metric 
to construct explicit examples of non-degenerate
5-dimensional metrics which satisfy the Legendre invariance condition.

\section{Systems with two degrees of freedom}
\label{sec:2dof}
\subsection{Two-dimensional geometrothermodynamics}
\label{sec:2gtd}

Let us consider a mono-component thermodynamic system with 
only two degrees of freedom. 
The thermodynamic 
phase space ${\mathcal T}$ is therefore a 5-dimensional manifold for which 
we choose the coordinates
\be
\{Z^A\} = \{\Phi, E^a, I^a\}=\{U,S,V,T,-P\}    \ ,
\ee
in accordance with standard notations of
thermodynamics \cite{callen}. The Gibbs 1-form 
\be
\Theta_{_G} = dU- TdS + P dV 
\ee
defines the contact structure of ${\mathcal T}$. The 2-dimensional 
space of
equilibrium states ${\cal E}$ is determined through the embedding mapping 
\be
\varphi :  (S,V) \longmapsto [U(S,V), S, V, T(S,V), P(S,V)] \ ,
\label{map2dof}
\ee 
so that $\varphi^* (\Theta_{_G}) =0$ implies the first law
of thermodynamics $dU = TdS - P dV$ with
\be
\frac{\partial U}{\partial S} = T\ ,\quad
\frac{\partial U}{\partial V} = - P \ .
\label{flaw2dof}
\ee
Notice that the fundamental equation $U=U(S,V)$ is contained in the definition of
the mapping (\ref{map2dof}), as mentioned in Section \ref{sec:contact}. 
The dependence of the intensive variables $T$ and $P$ in terms of the extensive 
variables $S$ and $V$ is induced through the conditions (\ref{flaw2dof}).  
The Euler identity and Gibbs-Duhem equation in this case can be written as 
\be
\beta U - TS + PV =0 \ ,\quad
S d T - V dP + (1-\beta) d U = 0 \ ,
\label{egd}
\ee
and reduce to the usual form when $\beta =1$, i.e., when $S$ and $V$ are
strictly extensive variables.  

A total Legendre transformation $
(U, S, V, T, - P) \longmapsto (\tilde U,\tilde S,\tilde V,\tilde T, - \tilde P)$
is defined in this case by means of the relationships
\be 
 U = \tilde U - \tilde S\tilde T +\tilde P\tilde V \ , \quad
 S = -\tilde T \ ,\ 
 V =\tilde P \ ,\ 
 T =\tilde S \ , \
 P = -\tilde V \ ,
\label{lt2dof}
\ee
and the corresponding transformation matrix reduces to 
\be
\frac{\partial Z^A}{\partial \tilde Z ^B}
=\left( \begin{array}{ccccc} 
\ 1 \ & \ - \tilde T\   & \ \ \tilde P\   & \ \ - \tilde S\  & \ \ \tilde V\   \\
0 & 0 & 0 & -1 & 0 \\
0 & 0 & 0 &  0 & 1 \\
0 & 1 & 0 &  0 & 0 \\
0 & 0 &-1 & 0 & 0 \\
\end{array}
\right) \, \,\, .
\label{mat2dof}
\ee
When compared to the general transformation matrix (\ref{matgen}), this matrix
presents several changes in the signs which are due
to the choice of the pressure as $P = - Z^4$. Also, it is convenient to write
the general 5-dimensional metric of ${\mathcal T}$ as $(I,J,... = 1,...4)$
\be
G = G_{00} d U^2 + 2 G_{0I} d U  d X^I + G_{IJ} d X ^I d X ^J \ ,\quad
 X^I =(S,V,T,P)\ ,
\label{met5d}
\ee
so that the negative sign of $Z^4$ does not affect the sign of the metric components.

According to our construction of geometrothermodynamics, we must demand that
the metric (\ref{met5d}) be invariant with respect to Legendre transformations.
If the components $G_{AB}$ are given as functions of the coordinates of the 
thermodynamic phase space, a total Legendre transformation leads to a set of algebraic
equations which are explicitly given in Appendix \ref{app:lt2dof}, where we also show the
existence of solutions. For a general 5-dimensional metric $G$ we can also
compute the corresponding 2-dimensional metric $g$ by using the pullback
of the mapping 
(\ref{map2dof}), i.e.  
\be
g=\varphi^*(G)= g_{11} d S ^2 + 2 g_{12} dS d V + g_{22} d V^2 \ ,
\ee
\begin{flushleft}
\bea
g_{11}\ = & \ & G_{11} + 2 G_{01} T + G_{00} T^2 
           + G_{33} U_{SS}^2 + G_{44} U_{SV}^2     \nonumber \\
         & + & 2 U_{SS} (G_{13} + G_{03} T - G_{34} U_{SV}) 
            - 2 U_{SV} ( G_{14} +  G_{04} T) \ , \label{g11}\\ 
g_{12}\ = & \ & G_{12} - P(G_{01}  + G_{00} T) + G_{02} T - G_{34} U_{SV}^2 \nonumber \\
         & + &  U_{SV} (G_{13} + G_{03} T - G_{24} + G_{04}P 
         + G_{33} U_{SS} + G_{44} U_{VV} ) \nonumber \\ 
         & + &  U_{SS}(G_{23} - G_{03}P - G_{34} U_{VV} )
             -  U_{VV} ( G_{14} + G_{04} T) \ , \label{g12} \\
g_{22}\ = & \ & G_{22} - 2 G_{02} P + G_{00} P^2 
           + G_{33} U_{SV}^2 + G_{44} U_{VV}^2     \nonumber \\
         & + & 2 U_{VV} (-G_{24} + G_{04} P - G_{34} U_{SV}) 
           + 2 U_{SV} ( G_{23}  -  G_{03} P) \ , \label{g22} 
\eea
\end{flushleft}
where the subscript of $U$ represent partial derivatives, and we used
equations (\ref{flaw2dof}) in the form $U_S = T,\ U_V = - P$. 
This is the most
general metric that can exist 
on the space of equilibrium states ${\mathcal E}$. 
The further specification 
of the fundamental equation $U=U(S,V)$ fixes completely the structure of $g$
in terms of the components $G_{AB}$ which,  in general, become arbitrary 
functions of $S$ and $V$, due to the definition of the embedding mapping 
(\ref{map2dof}). 
This arbitrariness can be restricted by considering only Legendre invariant
metrics $G$. In fact, if we apply a total Legendre transformation on
the 5-dimensional metric (\ref{met5d}), the new metric $\tilde G = G(Z^A 
\rightarrow \tilde Z ^A)$ 
will be Legendre invariant if its components satisfy the conditions 
(\ref{lt5d}). Furthermore, if we start from a Legendre invariant metric 
$\tilde G$ and perform the pullback determined by means of the mapping
\be
\tilde \varphi (\tilde S, \tilde V)\longmapsto 
[\tilde U(\tilde S,\tilde V), \tilde S, \tilde V, 
\tilde T(\tilde S,\tilde V), \tilde P(\tilde S,\tilde V)] \ ,
\label{map2tilde}
\ee 
we will obtain a new metric $ \tilde g = \tilde  \varphi^* (\tilde G)$
whose components satisfy the relationship
\be
\tilde g _{ab} (\tilde S, \tilde V) = g^\prime _{ab}(\tilde S, \tilde V)
=g _{ab}(Z^A =\tilde Z ^A , G_{AB}=\tilde G _{AB}) \ ,
\ee
where $g_{ab}$ are given in Eqs.(\ref{g11})--(\ref{g22}). In other words,
if the starting metric $G$ is Legendre invariant, the resulting metric 
$g=\varphi^*(G)$ is also Legendre invariant and its components are given
again as in Eqs.(\ref{g11})--(\ref{g22}).

We found the explicit form of the most general  2-dimensional metric $g$ 
on ${\mathcal E}$, starting from a general 5-dimensional metric 
on ${\mathcal T}$. 
The set of algebraic equations (\ref{g11})--(\ref{g22}) also allows us 
to analyze the inverse problem, i.e. if we know the explicit components
$g_{ab}$ of a metric on ${\mathcal E}$, we can find from equations
(\ref{g11})--(\ref{g22}) the most general form
of the metric $G$ which generates $g$ by means of the pullback.
In this manner, we can determine whether a given metric $g$ can be
obtained from a Legendre invariant metric $G$.  

As a simple example
let us consider the 2-dimensional Euclidean metric 
\be
g^E = dS^2 + d V^2 \ .
\label{euc2}
\ee
Introducing these values into (\ref{g11})--(\ref{g22}), we see that
the generating metric can be written as
\be
G^E = G_{00}[dU^2 + (T dS - PdV)^2] + dS^2 + dV^2 + 2 G_{0I}d X^I \Theta_{_G} \ ,
\ee
where $X^I = (S,V,T,P)$ and $\Theta_{_G} = dU - TdS + PdV$ is the fundamental
Gibbs 1-form for systems with two degrees of freedom. The components $G_{0A}$
remain as arbitrary functions of the coordinates of ${\mathcal T}$. This is an 
example of the freedom which exists in the determination of 2-dimensional metrics
on ${\mathcal E}$ in terms of the pullback of 5-dimensional metrics on 
${\mathcal T}$.
A straightforward calculation shows that the determinant
of $G^E$ vanishes identically, making this metric unsuitable for describing
a Riemannian structure on ${\mathcal T}$. Nevertheless, to show the details of
the procedure, we want to explore
the Legendre invariance of the Euclidean metric on ${\mathcal E}$. Applying
a total Legendre transformation (\ref{lt2dof})  to $G^E$, we obtain 
\bea
\tilde G ^E = & \ & 
G_{00}\tilde \Theta _{_G}^2  + d\tilde T ^2+ d \tilde P ^2 
\nonumber \\
& + & 2 (- G_{01} d\tilde T + G_{02} d \tilde P + G_{03} d \tilde S - G_{04} d \tilde V)
 \tilde \Theta _{_G} \ ,
\eea
where we used the Gibbs-Duhem relation $\tilde V d \tilde P - \tilde S d\tilde T =0$
in the new variables to simplify the final result.  Finally, we calculate the pullback
determined by the embedding mapping (\ref{map2tilde}) and obtain
\begin{flushleft}
\bea
\tilde g ^E & = & \tilde\varphi^* (\tilde G ^E) = \left(\frac{\partial^2 \tilde U}
{\partial \tilde E ^a \partial E ^b}\right)^2 d\tilde E ^a d\tilde E ^b 
 \nonumber \\ 
& = & (\tilde U _{\tilde S\tilde S}^2  + \tilde U _{\tilde S\tilde V}^2) d \tilde S ^2 
+ 2 \tilde U _{\tilde S\tilde V} (\tilde U_{\tilde S\tilde S} 
+ \tilde U_{\tilde V\tilde V}) d \tilde S d \tilde V 
+ (\tilde U _{\tilde V\tilde V}^2  + \tilde U _{\tilde S\tilde V}^2) d \tilde V ^2 \ ,
\label{weinq}
\eea
\end{flushleft} 
an expression which is functionally different from the starting metric (\ref{euc2}). 
This shows that the 2-dimensional Euclidean metric $g^E$ 
is not invariant with respect to Legendre transformations and, consequently, is 
not appropriate for describing any thermodynamic 
system in the context of geometrothermodynamics. 

\subsection{Weinhold's approach}
\label{sec:wei}

Weinhold proposed to study the geometry of thermodynamic systems by means
of a metric which is introduced {\it ad hoc} on the space of equilibrium states
${\mathcal E}$. In the case of a system with two degrees of freedom, it
corresponds to a 2-dimensional metric with components given 
in terms of the second derivatives of the internal energy with respect to 
the extensive variables. When the entropy representation is used, an alternative
metric can be defined again as the Hessian of the entropy (Ruppeiner's metric)
which, however, turns out to be conformally equivalent to Weinhold's metric. 
Since any metric on a 2-dimensional space is conformally equivalent to the 
Euclidean metric, it seems reasonable to expect a conformal equivalence between
all 2-dimensional metrics, when arbitrary diffeomorphisms are allowed. 

In the energy representation we are using here, Weinhold's metric is given by
\be
g^W = \left(\frac{\partial^2  U}
{\partial  E ^a \partial E ^b}\right) d E ^a d E ^b = U_{SS} d S^2 + 2 U_{SV} dS d V
+ U_{VV} d V^2 \ .
\label{wei}
\ee 
An important property of Weinhold's metric is that it is 
positive definite since its components  coincide with the matrix components 
of the second law of thermodynamics (\ref{slaw}). To determine the 5-dimensional
metric on ${\mathcal T}$ that generates Weinhold's metric, we introduce its 
components into equations (\ref{g11})--(\ref{g22}). After an algebraic rearrangement,
the resulting metric can be written as 
\be
G^W = G_{00} [ dU^2 -(T d S  -P d V)^2] + dS d V  - dT d P 
+ 2 G_{0I}d X^I \Theta_{_G} .
\ee
We see that the generating metric is not unique, but it possesses five degrees of
freedom which are contained in the five arbitrary functions $G_{0A}$. The rest 
of the components of $G_{AB}$ are fixed through the components of Weinhold's metric.
A special case of this generating metric was proposed in \cite{mru90} and corresponds
to the particular choice $G_{00}=1,\ G_{01}=-T,\ G_{02}= P,\ G_{03}=G_{04}=0$.

The next important question is whether the metric $G^W$ is invariant with respect
to arbitrary Legendre transformations. Introducing the explicit components of $G^W$
into the conditions of Legendre invariance given in Eqs.(\ref{lt5d}), one can 
see that they are not satisfied, independently of the values of the arbitrary
functions $G_{0A}$. Consequently, Weinhold's metric is not Legendre invariant,
making it inappropriate for describing thermodynamic systems, because the 
properties of the system would depend on the
thermodynamic potential used in $g^W$. This, of course, is a 
disadvantage which is sufficient for dismissing this metric as unphysical.
Nevertheless, to find the cause of the problem, we perform
a total Legendre transformation on $G^W$ to obtain a new metric $\tilde G ^W$.
Then we calculate its pullback 
$\tilde\varphi ^*(\tilde G ^W) = \tilde g ^W$ and obtain the metric
\be
\tilde g ^W = - \left(\frac{\partial^2 \tilde  U}
{\partial \tilde  E ^a \partial \tilde E ^b}\right) d \tilde E ^a d \tilde E ^b \ ,
\ee
which obviously contradicts the second law of thermodynamics (\ref{slaw}).
We see that the non-invariant character of this metric 
lies in the change of the metric signature. One could naively think that the 
additional ``parity transformation"  $\tilde U \rightarrow - \tilde U$ would
preserve invariance. However, this parity transformation on ${\mathcal E}$ 
is not allowed because it does not correspond to a Legendre transformation
on ${\mathcal T}$. Another possibility could be to consider a ``quadratic"
Weinhold's metric so that the signature becomes invariant. Unfortunately,
this simple alternative is also not allowed because the resulting metric 
is not invariant. In fact, this is the example we presented in Section 
\ref{sec:2gtd}. There, it was shown that the Euclidean metric $g^E$ turns
into the quadratic Weinhold's metric (\ref{weinq}) after a Legendre 
transformation. The inverse transformation would then transform the quadratic
Weinhold's metric into the Euclidean one which, as we mentioned above,
is not suitable for a geometrothermodynamic approach.

\subsection{Legendre invariant metrics}
\label{sec:lim}

Following the procedure described in Appendix \ref{app:lt2dof},
it is in principle very simple to find Legendre invariant metrics. 
By assuming a polynomial dependence of the metric components in terms
of the thermodynamic variables,  we derive in Eqs.(\ref{5dsol})
a particular solution which possesses twelve arbitrary real constants.   
Clearly, more general solutions can be obtained by starting from
 metric components with a different functional dependence.

In this Section, however, we will derive  several Legendre invariant
metrics by using a simple and intuitive method. From the definition
of a Legendre transformation given in (\ref{leg}), it follows
that metrics with no  $d\Phi$ terms and combinations of the form
$\sum_a [ (dE^a)^2 + (d I^a)^2]$ and 
$\sum_a [ I^a dE^a + E^a d I^a ]^2$ are invariant with respect to
Legendre transformations. This can be used as guide to construct
invariant metrics. In particular, for systems with two degrees
of freedom let us consider the following metric
\be
G^I= \Theta_{_G}^2 + c_1 T S d T d S + c_2 P V d P d V 
+ d S^2 + d V^2 + d T ^2 + d P^2 ,
\ee
which is Legendre invariant for any values of the real constants $c_1$ and $c_2$.
The term $\Theta_{_{_G}}^2$ has been added in order to avoid the degeneracy
of $G^I$. The pullback of this metric $\varphi^*(G^I) = g^I$ generates a 
2-dimensional metric on ${\mathcal E}$ which in a matrix 
representation can be written  as
\be
g^I = \mbox{\small{1}} \! \! 1 + (g^W)^2 +
\left(
\begin{array}{cc}
c_1 TS U_{SS} & \frac{1}{2}(c_1 TS - c_2 PV) U_{SV} \\
\frac{1}{2}(c_1 TS - c_2 PV) U_{SV} & - c_2 PV U_{VV}	
\end{array}
\right) \ ,
\label{gi}
\ee
where $ \mbox{\small{1}} \! \! 1$ is the unit matrix and $g^W$ represents 
Weinhold's metric (\ref{wei}). The special case $c_1=c_2=0$ corresponds
to Hern\'andez--Lacomba metric \cite{herlac98} which is also Legendre 
invariant. 

Consider now the following 5-dimensional non-degenerate metric
\be
G^{II} = \Theta_{_G}^2 + (TS - PV)( d S dT - d V d P) \ ,
\ee
which is also Legendre invariant. 
A straightforward computation of its pullback yields    
\be
g^{II} = U \left( U_{SS} d S^2 + 2 U_{SV} dS d V + U_{VV} d V^2\right) 
\label{gii}
\ee
where we used Euler's identity (\ref{egd}). We will see in the next Section
that this metric resembles the behavior of Weinhold's metric in the case of an 
ideal gas.

Clearly, the metrics presented above can easily be generalized to the case
of a system with $n$ degrees of freedom, preserving Legendre invariance. 
Dropping constant factors, the generalized $(2n+1)$-dimensional metrics
can be written as
\be
G^I = \Theta_{_G}^2 + \delta_{ab} E ^b I^a  d E^a d I ^b +
\delta_{ab} (d E^a d E ^b + d I ^a d I ^b ) \ ,
\ee
\be
G^{II} = \Theta_{_G}^2 + (\delta_{ab} E^a I^b)\ (\delta_{cd} d E^c d I^d)   \ ,
\label{genGII}
\ee
where $\Theta_{_G}$ is the Gibbs 1-form (\ref{gibbs}) 
for a thermodynamic system with $n$ degrees of freedom. 

All the examples we have presented here are characterized by the fact that
the metric components can be written as polynomials of the thermodynamic
variables. Solving the algebraic conditions described in Appendix 
\ref{app:lt2dof}, it is also possible to derive Legendre invariant metrics
of ${\mathcal T}$ whose components are rational functions of the variables.
The corresponding pullback generates metrics on ${\mathcal E}$ with rather
cumbersome structures which, nevertheless, are allowed in the context
of geometrothermodynamics.

\section{The ideal gas and thermodynamic length}
\label{sec:ig}

In our opinion, one of the most interesting properties of Weinhold's metric
is that its curvature vanishes in the case of a non-interacting thermodynamic
system, i.e. an ideal gas. For  more general systems, like the van der Waals
gas, the curvature
turns out to be non-vanishing and, therefore, it has been postulated that it 
could be used as a measure for thermodynamic interaction.  Although 
we have shown that Weinhold's metric is not compatible with the requirements of
geometrothermodynamics, we would like to preserve in our approach 
this interesting interpretation of curvature. To this end, we need to show
that there exists at least one Legendre invariant metric whose curvature
vanishes in the case of an ideal gas. We will show that in fact we already 
have a metric with this property.

Let us recall that for a 2-dimensional metric $g_{ab}$ there is only 
one independent component of the Riemann curvature  tensor $R_{abcd}$
which completely determines the curvature scalar 
$R= g^{ac} g^{bd} R_{abcd}$. In turn, the curvature scalar can be 
expressed as follows
\be
R =-\frac{1}{\sqrt{\det(g)}} \left[
\left(\frac{g_{11,2}-g_{12,1}}{\sqrt{\det(g)}}\right)_{,2} +
\left(\frac{g_{22,1}-g_{12,2}}{\sqrt{\det(g)}}\right)_{,1} \right] 
- \frac{1}{2\det(g)^2}\det(H) \ ,
\ee
with
\be
H=\left( \begin{array}{ccc} 
\ g_{11}  \ & \ g_{12} \   & \ g_{22} \ \\
\ g_{11,1}  \ & \ g_{12,1} \   & \ g_{22,1} \ \\
\ g_{11,2}  \ & \ g_{12,2} \   & \ g_{22,2} \ \\
\end{array}
\right) \, \,\, ,
\label{scaig}
\ee
where a comma denotes partial differentiation. 
As mentioned before, all the information about a specific system is
contained in the fundamental equation. In the case of an ideal gas
it can be written as \cite{burke}
\be
U(S,V)= V^{-2/3} \exp\left( \frac{2S}{3k} \right)    \ ,
\label{feig}
\ee
where $k$ is a constant. From this equation 
one can calculate the standard thermodynamic variables 
which are related through the equation of state $PV=kT$. Finally, the 
constant entropy  is also determined by means of  Euler's identity. 
Introducing this fundamental equation into the components of the metric $g^I$
given in (\ref{gi}) and computing the corresponding curvature scalar, we obtain
a rather cumbersome expression which is not a constant and
is always different
from zero, independently of the values of $c_1$ and $c_2$. 
We conclude that
the curvature of the metric $g^I$ is non zero for an ideal gas   and, consequently,
it  
cannot be used as a measure of thermodynamic
interaction. Consider now the metric $g^{II}$ as given in Eq.(\ref{gii}). 
A straightforward computation of its components with the fundamental equation 
(\ref{feig}) leads to the following result:
\be
g^{II}_{ig} = \frac{2U^2}{9k^2 V^2}\left( 2 V^2 dS^2 - 4 k V dS dV + 5 k ^2 dV^2\right)
 \ .
\ee
It is then easy to show that the corresponding 
scalar curvature vanishes identically for an ideal gas. Moreover, for more
general systems, like the van der Waals gas, the scalar curvature of $g^{II}$
does not vanish (see below). This is exactly the property we need in order 
to introduce the concept of thermodynamic length in an invariant manner. 

Let us consider a system with $n$ degrees of freedom whose phase space ${\cal T}$ 
is described by the general metric $G^{II}$ as given in (\ref{genGII}).
The computation of its pullback, using the conditions of thermodynamic
equilibrium (\ref{dual}) and Euler's identity (\ref{euler}), yields
\be
g^{II} = \Phi \left(\frac{\partial^2 \Phi}{\partial E^a \partial E^b}\right) d E^a d E^b \ .
\ee
This metric is defined on the space of equilibrium states ${\cal E}$ and, in
contrast to Weinhold's metric (\ref{wei}), is Legendre invariant due to 
the presence of the conformal factor $\Phi$. Since the curvature of this
metric is a measure of thermodynamic interaction, it seems plausible to use 
it also as a measure of ``length" on ${\cal E}$. 
If $t_1$ and $t_2$ represent two different
states on ${\cal E}$, we define their thermodynamic length as  
\be
L = \int_{t_1}^{t_2} \left[ 
\Phi\left( \frac{\partial^2 \Phi}{\partial E^a \partial E^b}\right) d E^a d E^b 
\right]^{1/2}\ .
\label{length}
\ee
Similar definitions have been proposed by Weinhold \cite{wei1} and 
Ruppeiner \cite{rup79}, using, however, non Legendre invariant metrics. 
We believe 
that our proposal (\ref{length}) will be useful for clarifying the
true significance of thermodynamic length.  
\section{The van der Waals gas and phase transitions}
\label{sec:vdw}
The fundamental equation for the van der Waals gas can be
written as
\be
U(S,V)= \frac   {e^{ \frac{2S}{3k}}} {(V-b)^{2/3}}   -\frac{a}{V} \ ,
\label{fevdw}
\ee
where $a$, $b$, and $k$ are constants. The thermodynamic interaction 
is determined through the constants $a$ and $b$. For this realistic gas
we find that the metric (\ref{gii}) reduces to 
\be
g^{II}_{vdW}= \frac{2}{9k^2} U(U+a/V)\left[ 2 dS^2 - \frac{4k }{V-b}dS dV
+\frac{5k^2 }{(V-b)^2}dV^2\right] -\frac{2aU}{V^3} dV^2
\ee
and the corresponding scalar curvature can be expressed as
\be
R=\frac{a{\cal P}(U,V,a,b)}{U^3(PV^3-aV+2ab)^2}
\ee
where ${\cal P}(U,V,a,b)$ is a polynomial which is always 
different from zero for any real values of $a$ and $b$.  In the limiting 
case $a=b=0$, the curvature vanishes and we turn back to the case of an ideal
gas. The curvature also vanishes for $a=0$ and $b\neq 0$, indicating that 
the constant $a$ is responsible for the non-ideal thermodynamic interactions,
whereas the constant $b$ plays a qualitative role in the description of 
 interactions \cite{san05c}.  When the condition 
\be 
PV^3-aV+2ab=0
\label{sta}
\ee
is satisfied, there are true curvature singularities 
which could be interpreted as showing the limit of applicability of 
geometrothermodynamics. Indeed, condition (\ref{sta}) for the van 
der Waals gas corresponds to the limit of thermodynamic stability 
\cite{callen} from which the information about critical points and
phase transitions can be obtained. 
Notice that condition (\ref{sta}) 
for curvature singularities represents the failure of local 
stability and, consequently, contains information about 
second order phase transitions only.  No information about first
order phase transitions can be obtained in this case from the
analysis of curvature.

The above  result can be formulated independently of the thermodynamic system
as follows. The expression for the scalar curvature (\ref{scaig}) shows 
that it is well defined only if $\det(g) > 0$ and becomes singular 
if $\det(g) =0$. In the case of the metric $g^{II}$ given in 
Eq.(\ref{gii}), these two conditions lead to 
\be 
U_{SS} U_{VV} - U_{SV}^2 \geq 0 \ ,
\ee
which corresponds to the convexity and stability conditions for
any thermodynamic system with two degrees of freedom in the
energy representation \cite{callen}.  
We conclude that the curvature 
singularities of the metric $g^{II}$ on ${\cal E}$ contain 
the information about the phase transitions of the thermodynamic
systems.

\section{Conclusions}
\label{sec:con}

In this work we formulated the fundamentals of geometrothermodynamics which
unifies the contact structure of the thermodynamic phase space 
${\mathcal T}$ with the metric structure of the space of thermodynamic 
equilibrium states ${\mathcal E}\subset {\mathcal T}$. 
The unifying object is a Riemannian metric structure on ${\mathcal T}$
which must be invariant with respect to Legendre transformations, and 
compatible with the metric structure induced on ${\mathcal E}$ by means
of the pullback. 
It was shown how the first and second laws of thermodynamics, as well
as Euler's identity and the Gibbs-Duhem relation, are invariant 
with respect to Legendre transformations. We found explicitly 
the conditions under which the Legendre transformations
are also isometries for a metric structure on ${\mathcal T}$. This
leads to a system of algebraic equations for the components of
the metric. Several particular solutions were found for this system.
All the solutions found turned out to be represented as polynomials
in terms of the coordinates of ${\mathcal T}$, but it is possible
to derive more general solutions. 

We analyzed in detail the case of thermodynamic systems with two 
degrees of freedom. Weinhold's approach was investigated for 
this case and we found that it does not satisfy the conditions
of Legendre invariance and, therefore, cannot be used to 
describe thermodynamic systems in an invariant way. In particular,
we saw that it is in conflict with the second law of thermodynamics. 
As explicit examples of simple thermodynamic systems we investigated
the ideal gas and van der Waals gas. Within the set of Legendre invariant 
metrics 
included in this work, we found one particular metric whose curvature 
vanishes in the case of an ideal gas, and is different from zero otherwise. 
This result makes this metric
especially suitable for measuring the interaction in thermodynamic 
systems. 

We proposed a definition of thermodynamic length which,
in contrast to other definitions known in the literature, is Legendre
invariant. Furthermore, we proved that in the case of thermodynamic
systems with two degrees of freedom, the curvature 
singularities of a specific Legendre invariant metric indicates
the existence of critical points and phase transitions.  

It would be interesting to continue the study of the metrics
presented here in the case of more complicated thermodynamic
systems like the Ising model, multi-component
ideal gas, etc. This would shed light on the significance of
these metrics. 
On the other hand, our approach opens the possibility
of investigating simultaneously the geometric structure not only 
in the space of equilibrium states, but also at the level of
the thermodynamic phase space. These tasks will be treated in future
works. 

Our geometrothermodynamic approach opens new possibilities to 
study thermodynamic systems in terms of geometric objects in an invariant manner.
At the same time, we will have to face new challenges. In particular,
we have found a huge arbitrariness in our description. For instance,
the study of the ideal gas allowed us to stand out the metric $g^{II}$
given in Section \ref{sec:ig}, because its curvature vanishes in 
the limiting case of an ideal gas. However,
our results seem to indicate that this is not a special property of
that metric. With the help of the algebraic conditions 
presented in Appendix \ref{app:lt2dof}, it would be possible 
to find more metrics with the same property. It seems that
we need additional criteria to select or classify metrics
on the thermodynamic phase space. Understanding the metric 
in this context represents an interesting problem for 
future investigations. 

\section*{Acknowledgements} 
This work was supported in part by Conacyt, M\'exico, grant 48601.



\appendix

\section{Legendre transformations  for systems with two degrees of freedom}
\label{app:lt2dof}

The components $G_{AB}$ of an arbitrary metric in a 5-dimensional thermodynamic phase
space transform under a Legendre transformation (\ref{leg}) 
according to the rule 
\be
G_{AB} \rightarrow \tilde G _{AB} = \frac{\partial Z^C}{\partial \tilde Z ^A}
  \frac{\partial Z^D}{\partial \tilde Z ^B} G_{CD} \ ,
\ee
where the transformation matrix is given in Eq.(\ref{mat2dof}). 
Using the representation (\ref{met5d}), a total Legendre transformation leads to
\begin{flushleft}
\bea
 \tilde G _{0 0}\  & = &\  G _{0 0}\ , \nonumber\\ 
 \tilde G _{0 1} \ & = & \  G _{0 3} -  G _{0 0} \tilde T \ , \nonumber\\ 
\tilde G _{0 2}\  & = & \  -  G _{0 4} +  G _{0 0} \tilde P \ , \nonumber\\ 
 \tilde G _{0 3}\ & = & \  - ( G _{0 1} +  G _{0 0} \tilde S )\ , \nonumber\\ 
 \tilde G _{0 4}\ & = & \  G _{0 2} +  G _{0 0} \tilde V \ , \nonumber\\ 
 \tilde G _{1 1}\ & = & \  G _{3 3} - 2  G _{0 3} \tilde T  +  G _{0 0} \tilde T ^2\ , \nonumber\\ 
 \tilde G _{1 2}\ & = & \  -  G _{3 4} +  G _{0 4} \tilde T  +  G _{0 3} \tilde P   -  G _{0 0} 
 \tilde P   \tilde T \ , \nonumber\\ 
 \tilde G _{1 3}\ & = & \  -  G _{1 3} -  G _{0 3} \tilde S  +  G _{0 1} \tilde T  +  G _{0 0} \tilde S  \tilde T \ , \\ 
\tilde G _{1 4}\  & = & \  G _{2 3} +  G _{0 3} \tilde V  -  G _{0 2} \tilde T  -  G _{0 0} \tilde T  \tilde V \ , \nonumber\\ 
 \tilde G _{2 2}\ & = & \  G _{4 4} - 2  G _{0 4} \tilde P   +  G _{0 0} \tilde P  ^2\ , \nonumber\\ 
 \tilde G _{2 3}\ & = & \  G _{1 4} +  G _{0 4} \tilde S  -  G _{0 1} \tilde P   -  G _{0 0} \tilde P   \tilde S \ , \nonumber\\ 
 \tilde G _{2 4}\ & = & \  -  G _{2 4} -  G _{0 4} \tilde V  +  G _{0 2} \tilde P   +  G _{0 0} \tilde P  \tilde V  \ , \nonumber\\ 
 \tilde G _{3 3}\ & = & \  G _{1 1} + 2  G _{0 1} \tilde S  +  G _{0 0} \tilde S ^2\ , \nonumber\\ 
 \tilde G _{3 4}\ & = & \  - ( G _{1 2} +  G _{0 2} \tilde S  +  G _{0 1} \tilde V   +  G _{0 0} \tilde S  \tilde V  )\ , \nonumber\\ 
 \tilde G _{4 4}\ & = & \  G _{2 2} + 2  G _{0 2} \tilde V   +  G _{0 0} \tilde V  ^2\ . \nonumber
 \label{lt5d}
  \eea
\end{flushleft}
This represents a system of algebraic equations for the components $G_{AB}$
which, if fulfilled, guarantees Legendre invariance. The explicit 
dependence of the above system from the coordinates of the thermodynamic 
phase space points to a particular solution which can be derived by
considering an ansatz where $G_{0A}$ depends linearly on $Z^A$ and the $G_{IK}$'s
are quadratic functions of the coordinates. A lengthly but straightforward
computation shows that the metric
\begin{flushleft}
\bea
G_{00}\ & = & \ c_1 \ , \nonumber \\
G_{01}\ & = & \ c_2 T  \ , \nonumber \\
G_{02}\ & = & \ - c_2 P \ , \nonumber \\
G_{03}\ & = & -(c_1+c_2) S   \ , \nonumber \\
G_{04}\ & = & (c_1+c_2) V  \ , \nonumber \\
G_{11}\ & = & c_3 T^ 2 +c_4   \ , \nonumber \\
G_{12}\ & = & - c_5 T P +c_6   \ , \nonumber \\
G_{13}\ & = & c_7 S T    \ ,  \\
G_{14}\ & = & - c_8 V T  +c_9   \ , \nonumber \\
G_{22}\ & = & c_{10}  P ^2 +c_{11}   \ , \nonumber \\
G_{23}\ & = & -c_{8}  P S - c_{9}   \ , \nonumber \\
G_{24}\ & = & c_{12}V   P   \ , \nonumber \\
G_{33}\ & = & ( c_1 + 2 c_2 + c_3)S^2   +c_{4}   \ , \nonumber \\
G_{34}\ & = & -( c_1 + 2 c_2 + c_5)S V    - c_{6}   \ , \nonumber \\
G_{44}\ & = & ( c_1 + 2 c_2 + c_{10})V^2   +c_{11}   \ , \nonumber
\label{5dsol}
\eea
\end{flushleft}
is completely invariant with respect to total Legendre transformations.  
Here $c_1, ..., c_{12}$ represent arbitrary real constants and in 
general ${\rm det}(G_{AB})\neq 0$. This particular solution  shows
that there exist non-trivial thermodynamic metrics which can be used to 
introduce Riemannian structures in the thermodynamic phase space. 
It is possible to derive other particular solutions by assuming 
a polynomial dependence of higher degree for all metric 
components $G_{AB}$. We also investigated the case of rational and 
exponential functions, and obtained similar results.




\end{document}